\documentclass[dvips]{acta}
\usepackage{supertabular,lscape,epsfig}
\usepackage{amssymb}
\usepackage{amsmath}
\DeclareSymbolFont{ppa}{OT1}{ppl}{m}{it}
\DeclareMathSymbol{\vv}{\mathalpha}{ppa}{'166}

\newfont{\hb}{rphvb at 10pt}
\newfont{\hbo}{rphvbo at 10pt}
\newfont{\bitt}{rptmbi at 12pt}
\newfont{\bits}{rptmbi at 11pt}

\SetPages{1}{1}

\SetVol{65}{2015}

\begin{document}

\newcommand{\TabCapp}[2]{\begin{center}\parbox[t]{#1}{\centerline{
  \small {\spaceskip 2pt plus 1pt minus 1pt T a b l e}
  \refstepcounter{table}\thetable}
  \vskip2mm
  \centerline{\footnotesize #2}}
  \vskip3mm
\end{center}}

\newcommand{\TTabCap}[3]{\begin{center}\parbox[t]{#1}{\centerline{
  \small {\spaceskip 2pt plus 1pt minus 1pt T a b l e}
  \refstepcounter{table}\thetable}
  \vskip2mm
  \centerline{\footnotesize #2}
  \centerline{\footnotesize #3}}
  \vskip1mm
\end{center}}

\newcommand{\MakeTableSepp}[4]{\begin{table}[p]\TabCapp{#2}{#3}
  \begin{center} \TableFont \begin{tabular}{#1} #4
  \end{tabular}\end{center}\end{table}}

\newcommand{\MakeTableee}[4]{\begin{table}[htb]\TabCapp{#2}{#3}
  \begin{center} \TableFont \begin{tabular}{#1} #4
  \end{tabular}\end{center}\end{table}}

\newcommand{\MakeTablee}[5]{\begin{table}[htb]\TTabCap{#2}{#3}{#4}
  \begin{center} \TableFont \begin{tabular}{#1} #5
  \end{tabular}\end{center}\end{table}}

\newfont{\bb}{ptmbi8t at 12pt}
\newfont{\bbb}{cmbxti10}
\newfont{\bbbb}{cmbxti10 at 9pt}
\newcommand{\uprule}{\rule{0pt}{2.5ex}}
\newcommand{\douprule}{\rule[-2ex]{0pt}{4.5ex}}
\newcommand{\dorule}{\rule[-2ex]{0pt}{2ex}}
\def\thefootnote{\fnsymbol{footnote}}
\begin{Titlepage}
\Title{The OGLE Collection of Variable Stars.\\
Classical Cepheids in the Magellanic System\footnote{Based on observations
obtained with the 1.3-m Warsaw telescope at the Las Campanas Observatory of
the Carnegie Institution for Science.}}
\Author{I.~~S~o~s~z~y~ñ~s~k~i$^1$,~~
A.~~U~d~a~l~s~k~i$^1$,~~
M.\,K.~~S~z~y~m~a~ñ~s~k~i$^1$,~~
D.~~S~k~o~w~r~o~n$^1$,\\
G.~~P~i~e~t~r~z~y~ñ~s~k~i$^1$,~~
R.~~P~o~l~e~s~k~i$^{1,2}$,~~
P.~~P~i~e~t~r~u~k~o~w~i~c~z$^1$,~~
J.~~S~k~o~w~r~o~n$^1$,\\
P.~~M~r~ó~z$^1$,~~
S.~~K~o~z~³~o~w~s~k~i$^1$,~~
\L.~~W~y~r~z~y~k~o~w~s~k~i$^1$,\\
K.~~U~l~a~c~z~y~k$^3$~~
and~~M.~~P~a~w~l~a~k$^1$}
{$^1$Warsaw University Observatory, Al.~Ujazdowskie~4, 00-478~Warszawa, Poland\\
e-mail: (soszynsk,udalski)@astrouw.edu.pl\\
$^2$Department of Astronomy, Ohio State University, 140 W. 18th Ave., Columbus, OH~43210, USA\\
$^3$Department of Physics, University of Warwick, Gibbet Hill Road,\\ Coventry, CV4 7AL, UK}
\Received{~}
\end{Titlepage}

\Abstract{We present here a nearly complete census of classical Cepheids in
the Magellanic System. The sample extends the set of Cepheids published
in the past by the Optical Gravitational Lensing Experiment (OGLE) to the
outer regions of the Large (LMC) and Small Magellanic Cloud (SMC). The
entire collection consists of 9535 Cepheids of which 4620 belong to the
LMC and 4915 are members of the SMC. We provide the {\it I}- and {\it
  V}-band time-series photometry of the published Cepheids, their finding
charts, and basic observational parameters.

Based on this unique OGLE sample of Cepheids we present updated
period--luminosity relations for fundamental, first, and second mode of
pulsations in the {\it I-} and {\it V}-bands and for the $W_I$
extinction-free Wesenheit index. We also show the distribution of classical
Cepheids in the Magellanic System.

The OGLE collection contains several classical Cepheids in the Magellanic
Bridge -- the region of interaction between the Magellanic Clouds. The
discovery of classical Cepheids and their estimated ages confirm the
presence of young stellar population between these galaxies.}{Cepheids -- Magellanic Clouds -- Catalogs}

\Section{Introduction}
Classical Cepheids (also known as $\delta$~Cephei stars or type~I Cepheids)
are objects of particular interest for a variety of reasons. They are
primary distance indicators within the Milky Way and to extragalactic
systems up to the Virgo cluster. These stars are also useful tracers of the
young stellar population in our and other galaxies. Furthermore, classical
Cepheids are excellent objects for constraining stellar pulsation and
evolution models.

Cepheids in the Large (LMC) and Small Magellanic Cloud (SMC) played a
special historical role, since the famous period--luminosity (PL) relation
(also called the Leavitt law) was first noticed in these galaxies (Leavitt
1908). Nowadays, the LMC and SMC are also important targets for studying
$\delta$~Cep stars because both galaxies contain the largest known
collection of these pulsators among all stellar environments, including the
Milky Way. The Optical Gravitational Lensing Experiment (OGLE) has
regularly published large samples of classical Cepheids in the Magellanic
Clouds discovered in consecutive phases of the survey since 1999 (\eg
Udalski \etal 1999ab).

The latest samples of the OGLE classical Cepheids consisted of 3375 of
these stars in the LMC (Soszyñski \etal 2008) and 4630 in the SMC
(Soszyñski \etal 2010). These releases were followed by an avalanche of
papers presenting various applications of the OGLE Cepheid data. The PL
relations in a wide range of photometric passbands, from optical to
infrared, were derived and studied based on the OGLE Cepheids (\eg Ngeow
\etal 2009, 2010, 2012, Majaess \etal 2011, Storm \etal 2011, Ripepi \etal
2012, Macri \etal 2015), which led to precise measurements of the distances
to the Magellanic Clouds (\eg Inno \etal 2013, Scowcroft \etal 2016, Ngeow
\etal 2015) and three-dimensional geometric models of both galaxies (\eg
Haschke \etal 2012ab, Moretti \etal 2014, Subramanian and Subramaniam 2015,
Scowcroft \etal 2016). The morphology of the OGLE light curves of classical
Cepheids were a subject of intensive comparative studies (\eg Deb and
Singh 2009, Pejcha and Kochanek 2012, Klagyivik \etal 2013, Bhardwaj \etal
2015). The OGLE Cepheids were used as training sets for automatic
classification systems (\eg Deb and Singh 2009, Long \etal 2012, Kim \etal
2014) and as a basis of the asteroseismic considerations (\eg Dziembowski
and Smolec 2009, Moskalik and Ko³aczkowski 2009, Smolec and Moskalik 2010,
Dziembowski 2012). Cepheids in eclipsing binary systems discovered by OGLE
allowed finding the solution of the mass discrepancy problem of Cepheids
(\eg Pietrzyñski \etal 2010, 2011, Gieren \etal 2014, Pilecki \etal 2015).

In this paper, we extend the OGLE-III samples of classical Cepheids in the
Magellanic Clouds (Soszyñski \etal 2008, 2010) by adding variables detected
in the OGLE-IV fields covering the outskirts of both galaxies and the
region between them, the so called Magellanic Bridge. The entire collection
constitutes a nearly complete census of $\delta$~Cep stars in the
Magellanic Clouds. We provide OGLE-IV light curves of the newly detected
and previously known Cepheids, increasing the time span of the OGLE
observations to over 18 years in the central regions of both galaxies.

The article is organized as follows. Section~2 briefly describes the
OGLE-IV photometric data used in this investigation. Section~3 provides
details on the Cepheid identification and classification. In Section~4, we
show the OGLE Cepheid collection itself. In Section~5, we estimate the
completeness of our sample and compare it with other sets of Cepheids in
the Magellanic Clouds. In Section~6, we present updated PL relations for
fundamental, first-, and second-overtone pulsation modes in the $W_I$
Wesenheit index and {\it VI} photometric bands. We also report there the
discovery of several classical Cepheids in the Magellanic Bridge and
discuss its consequences. The conclusions are summarized in Section~7.

\Section{Observations and Data Reduction}
This study is based on the {\it I}- and {\it V}-band time-series photometry
collected during the fourth phase of the OGLE survey (OGLE-IV) between
March 2010 and July 2015. The project has been conducted with the 1.3-m
Warsaw telescope at Las Campanas Observatory in Chile (the observatory is
operated by the Carnegie Institution for Science). The telescope is
equipped with a 32-CCD detector mosaic camera, covering approximately
1.4~square degrees on the sky with the scale of 0.26~arcsec/pixel. OGLE-IV
is monitoring a total of about 650~square degrees in the LMC, SMC, and in
the Magellanic Bridge which links the two galaxies. The number of point
sources in the OGLE-IV databases is about 58 million in the LMC, 12 million
in the SMC, and 5 million in the Magellanic Bridge.

Most of the observations (from about 100 to over 750, depending on the
field) were secured through the Cousins {\it I}-band filter, the remaining
data points (from several to 260) were obtained in the Johnson {\it
V}-band. Data reduction of the OGLE images was performed using the standard
OGLE photometric data pipeline (Udalski \etal 2015), based on the
Difference Image Analysis technique (Alard and Lupton 1998, Wo¼niak
2000). Detailed description of the instrumentation, photometric reductions
and astrometric calibrations of the OGLE observations is provided in
Udalski \etal (2015).

\Section{Selection and Classification of Classical Cepheids}
An extensive search for variable stars was preceded by a period search
performed for all point sources observed by the OGLE-IV survey in the
Magellanic Clouds. The Fourier periodogram was calculated for each {\it
I}-band light curve with at least 30 observing points. We used the {\sc
Fnpeaks} code\footnote{
http://helas.astro.uni.wroc.pl/deliverables.php?lang=en\&active=fnpeaks}
written by Z.~Ko³aczkowski. Having the primary periods, their
signal-to-noise ratios, amplitudes of variability, mean magnitudes, and
parameters of the Fourier decomposition of the light curves we conducted a
semi-manual search for variable stars. We visually inspected the light
curves with the largest signal-to-noise ratios and stars located within a
wide strip in the PL diagram covering all types of Cepheids and RR~Lyrae
type stars. The detected variable stars were initially classified as
pulsators, eclipsing binaries, and other variables.

The first group was then divided into classical Cepheids, anomalous
Cepheids, type II Cepheids, RR~Lyrae stars, $\delta$~Scuti stars, and
long-period variables. The most important criteria used in this
classification were the light curve shapes quantified by their Fourier
parameters, position of the stars in the PL diagrams, and the period ratios
(for multi-periodic variables). For example, Soszyñski \etal (2015) showed
that Fourier parameters $\phi_{21}$ and $\phi_{31}$ are useful tools to
distinguish between classical and anomalous Cepheids. Because $\delta$~Sct
stars and overtone classical Cepheids form a continuous relationship in the
PL diagram, we adopted a boundary period of 0.23~d to separate both groups.
In ambiguous cases we carefully inspected the light curves, checked other
properties of the stars, before deciding on the final classification.
Nevertheless, one should be aware that for a limited number of individual
objects our classification may be incorrect. We mark doubtful stars in the
remark file of the catalog.

\MakeTable{
l@{\hspace{8pt}} c@{\hspace{6pt}} | l@{\hspace{8pt}} c@{\hspace{6pt}}}{12.5cm}
{Reclassified stars from the OGLE-III catalogs of classical Cepheids in the Magellanic Clouds}
{\hline \noalign{\vskip3pt}
\multicolumn{1}{c}{Identifier} & New            & \multicolumn{1}{c}{Identifier} & New            \\
                       & classification &                        & classification \\
\noalign{\vskip3pt}
\hline
\noalign{\vskip3pt}
OGLE-LMC-CEP-0320 & RR Lyr        & OGLE-SMC-CEP-1636 & Anom. Cepheid  \\
OGLE-LMC-CEP-0665 & Other         & OGLE-SMC-CEP-1650 & Anom. Cepheid  \\
OGLE-LMC-CEP-1277 & RR Lyr (RRd)  & OGLE-SMC-CEP-1826 & Anom. Cepheid \\
OGLE-LMC-CEP-3063 & Eclipsing     & OGLE-SMC-CEP-2089 & Anom. Cepheid \\
OGLE-SMC-CEP-0008 & Anom. Cepheid & OGLE-SMC-CEP-2169 & Anom. Cepheid \\
OGLE-SMC-CEP-0080 & Anom. Cepheid & OGLE-SMC-CEP-2210 & Anom. Cepheid \\
OGLE-SMC-CEP-0167 & Anom. Cepheid & OGLE-SMC-CEP-2343 & Anom. Cepheid \\
OGLE-SMC-CEP-0174 & Anom. Cepheid & OGLE-SMC-CEP-2485 & Anom. Cepheid \\
OGLE-SMC-CEP-0220 & Anom. Cepheid & OGLE-SMC-CEP-2659 & Anom. Cepheid \\
OGLE-SMC-CEP-0252 & Anom. Cepheid & OGLE-SMC-CEP-2714 & Anom. Cepheid \\
OGLE-SMC-CEP-0269 & Anom. Cepheid & OGLE-SMC-CEP-2740 & Anom. Cepheid \\
OGLE-SMC-CEP-0326 & Anom. Cepheid & OGLE-SMC-CEP-2834 & RR Lyr        \\
OGLE-SMC-CEP-0354 & Anom. Cepheid & OGLE-SMC-CEP-2862 & Anom. Cepheid \\
OGLE-SMC-CEP-0366 & Anom. Cepheid & OGLE-SMC-CEP-3136 & Anom. Cepheid \\
OGLE-SMC-CEP-0475 & Anom. Cepheid & OGLE-SMC-CEP-3540 & Anom. Cepheid \\
OGLE-SMC-CEP-0532 & Anom. Cepheid & OGLE-SMC-CEP-3698 & Anom. Cepheid \\
OGLE-SMC-CEP-0677 & Anom. Cepheid & OGLE-SMC-CEP-3814 & Anom. Cepheid \\
OGLE-SMC-CEP-1078 & Anom. Cepheid & OGLE-SMC-CEP-3957 & Anom. Cepheid \\
OGLE-SMC-CEP-1082 & Anom. Cepheid & OGLE-SMC-CEP-4369 & Anom. Cepheid \\
OGLE-SMC-CEP-1129 & Anom. Cepheid & OGLE-SMC-CEP-4391 & Anom. Cepheid \\
OGLE-SMC-CEP-1241 & Anom. Cepheid & OGLE-SMC-CEP-4582 & Anom. Cepheid \\
OGLE-SMC-CEP-1355 & Anom. Cepheid & OGLE-SMC-CEP-4608 & Anom. Cepheid \\
OGLE-SMC-CEP-1476 & Anom. Cepheid & OGLE-SMC-CEP-4621 & Anom. Cepheid \\
\noalign{\vskip3pt}
\hline}
We also examined OGLE-IV light curves of Cepheids discovered during the
OGLE-II and OGLE-III phases of the project and cataloged by Soszyñski \etal
(2008, 2010). Forty six objects were reclassified as other types of
variable stars and they were removed from the OGLE collection of classical
Cepheids in the Magellanic System. A list of these stars is given in
Table~1. The vast majority of them are variables in the SMC which were
reclassified by Soszyñski \etal (2015) as anomalous Cepheids. Classical
and anomalous Cepheids in the SMC share very similar light curve morphology
that can be associated with the low metallicity of the stars in this
galaxy. Further several candidates for classical Cepheids in the OGLE-III
collection have very dubious classification, but they were left on the
list, since we cannot rule out the possibility that they are real Cepheids
with unusual properties.

\begin{figure}[t]
\centerline{\includegraphics[width=13.4cm]{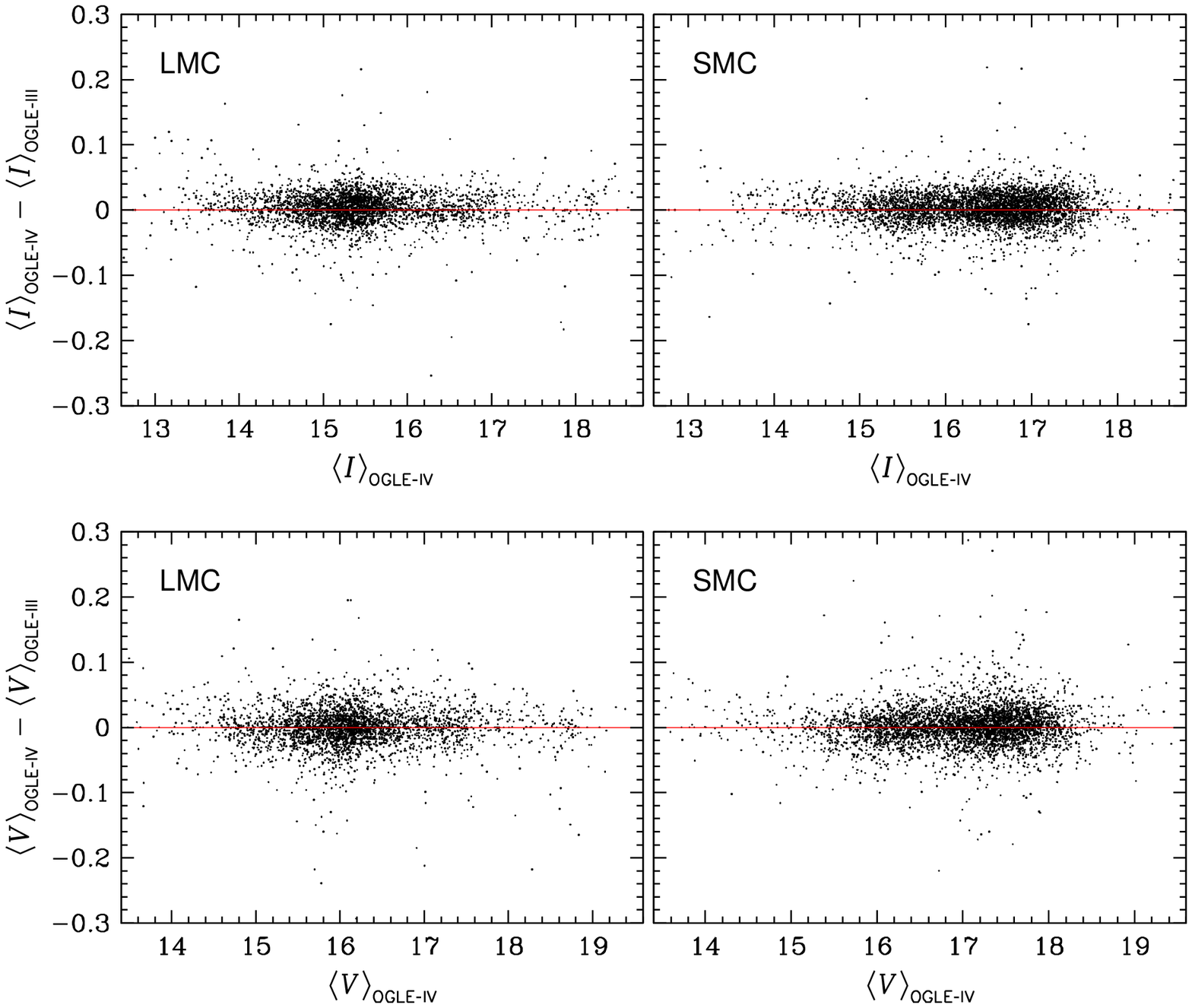}}
\vskip-6.4cm
\FigCap{Comparison between OGLE-III and OGLE-IV {\it I}-band ({\it upper
panels}) and {\it V}-band ({\it lower panels}) mean magnitudes of classical
Cepheids in the LMC ({\it left panels}) and SMC ({\it right panels}).}
\end{figure}

Periods, mean magnitudes, amplitudes, and other parameters of the Cepheids
known from the previous stages of the OGLE project were recalculated using
the OGLE-IV light curves accumulated in the years 2010--2015. The OGLE-III
and OGLE-IV photometry were obtained with different instrumental
configurations, in particular with different filters and CCD detectors, but
both datasets were transformed to the standard photometric system with the
systematic uncertainties of the calibration zero point up to 0.02~mag. In
Fig.~1, we compare the OGLE-III and OGLE-IV mean magnitudes of Cepheids in
the {\it I}- and {\it V}-filters. The agreement between both data sets is
very good. The mean difference between magnitudes measured from the OGLE-IV
and OGLE-III photometry is smaller than 0.003~mag in both filters and in
both galaxies. The standard deviation is smaller than 0.03~mag. Outliers
in Fig.~1 can be explained by crowding and blending by unresolved stars.

\vspace*{7pt}
\Section{Classical Cepheids in the Magellanic Clouds} 
\vspace*{5pt}
The OGLE collection of classical Cepheids in the Magellanic Clouds
comprises 9535 objects (4620 in the LMC and 4915 in the SMC), of which 5168
pulsate solely in the fundamental mode (F), 3530 are single-mode
first-overtone pulsators (1O), 117 oscillate purely in the second overtone
(2O), 711 stars are double-mode Cepheids, and nine -- triple-mode
Cepheids. The data on all these objects are available through the OGLE
anonymous FTP sites or {\it via} the OGLE web interface:
\begin{center}
{\it ftp://ftp.astrouw.edu.pl/ogle/ogle4/OCVS/lmc/cep/}\\
{\it ftp://ftp.astrouw.edu.pl/ogle/ogle4/OCVS/smc/cep/}\\
{\it http://ogle.astrouw.edu.pl}
\end{center}

OGLE FTP sites are organized as follows. The files named {\sf ident.dat}
contain the full lists of classical Cepheids in both galaxies. For each
star we provide its identifier, J2000 equatorial coordinates, mode(s) of
pulsation, fields and internal numbers in the OGLE-IV, OGLE-III, and
OGLE-II photometric databases (if available) and the cross-matches with the
extragalactic part of the General Catalogue of Variable Stars (GCVS,
Artyukhina \etal 1995). The identifiers of the Cepheids are the same as in
the OGLE-III catalogs -- OGLE-LMC-CEP-NNNN and OGLE-SMC-CEP-NNNN -- where
NNNN is a four-digit number. The newly detected Cepheids have numbers
larger than 3375 and 4630 in the LMC and SMC, respectively, and are
organized by increasing right ascension.

Files {\sf cep*.dat} contain observational parameters of various types of
classical Cepheids: their mean magnitudes in the {\it I}- and {\it
V}-bands, pulsation periods, epochs of the maximum light, peak-to-peak
{\it I}-band amplitudes, and Fourier parameters derived from the {\it
I}-band light curves. The pulsation periods were refined with the {\sc
Tatry} code (Schwarzenberg-Czerny 1996) using OGLE-IV observations
obtained between 2010 and 2015. To study the long-term behavior of 
Cepheids in the Magellanic Clouds, we recommend to merge the OGLE-IV light
curves with the photometry obtained during the previous phases of the OGLE
survey and published by Soszyñski \etal (2008, 2010). One should take into
account that smaller or larger differences between mean magnitudes and
amplitudes for individual objects are possible (see Section~3). A limited
number of Cepheids published by Soszyñski \etal (2008, 2010) do not have
OGLE-IV measurements, because they fell into the gaps between the CCD
detectors of the OGLE-IV mosaic camera. For these stars we provide their
parameters from the OGLE-III Cepheid data release.

Files containing OGLE-IV time-series photometry in the {\it I}- and {\it
V}-bands are stored in the directory {\sf phot}. Finding charts can be
found in the directory {\sf fcharts}. These are $60\arcs\times60\arcs$
subframes of the {\it I}-band reference images, oriented with North at
the top and East to the left. Additional information on some Cepheids
(uncertain classification, eclipsing or ellipsoidal variability,
secondary periods, etc.) are given in the file {\sf remarks.txt}.

Most of the classical Cepheids detected in the Magellanic Bridge (see
Section~6.3) were included in the SMC sample, since it is widely believed
that gas present in the Bridge was drawn out of the SMC through tidal
forces during a close encounter of the two galaxies, which took place about
250~Myr ago (Mathewson 1985, Muller \etal 2004).

\vspace*{7pt}
\Section{Completeness of the Sample}
\vspace*{5pt}
The completeness of the OGLE Collection of classical Cepheids in the
Magellanic System is the highest in the area covered by the OGLE-II and
OGLE-III fields because these regions were independently searched for
variable stars in the past (Udalski \etal 1999ab, Soszyñski \etal 2008,
2010). The central 40 square degrees in the LMC and 14 square degrees in
the SMC contain most of the classical Cepheids in these galaxies. The
completeness of the OGLE sample outside these regions is limited by the
gaps between the OGLE-IV fields and between the CCD detectors of the mosaic
camera. We estimate that currently about 7\% of stars may be missed in the
outer regions of the Clouds due to the gaps in our coverage. It has to be,
however, noted that we plan to conduct additional search for classical
Cepheids in these currently uncovered area using a new set of reference
images which practically fully fill all the gaps between CCD
detectors. After this minor update planned for 2016, the OGLE collection of
classical Cepheids should contain practically all classical Cepheids from
the Magellanic System.

The completeness of the sample in the area covered by the OGLE observations
may be calculated based on the stars with double entries in the
database. Neighboring OGLE-IV fields overlap (\cf Udalski \etal 2015) and
some Cepheids have been detected twice, independently in both
fields. Please note, however, that the final version of our collection
contains only one light curve per star -- usually the one with larger
number of epochs.

We found that 897 classical Cepheids are located in the overlapping regions
of the neighboring fields. Taking into account light curves with at least
100 observing points, we obtained 854 pairs, so we had a chance to identify
1708 counterparts. We independently detected 1698 of them, which gives the
general completeness of our sample larger than 99\%. Six of the missing
Cepheids have nearly sinusoidal light curves and were initially classified
as ellipsoidal variables, two objects have exceptionally large noise due to
the proximity of bright stars, and two Cepheids have periods close to 1 or
2~days which affected the Fourier series fits to their light curves.

We also compared our collection with other samples of Cepheids in the
Magellanic Clouds. The MACHO catalog of Cepheids in the LMC (Alcock
\etal 1999) consists of 1800 objects. Our collection does not contain 11
of these stars, of which three are located beyond the OGLE fields and
eight are misclassified by the MACHO project non-pulsating stars:
eclipsing or spotted variables.

Kim \etal (2014) released a list of 117\,234 candidates for periodic
variable stars in the LMC based on EROS-2 microlensing survey
observations. The sources were categorized into several variability types
using an automatic random forest method. The EROS-2 set contains variable
stars not observed by the previous stages of the OGLE survey, of which 638
sources were classified by Kim \etal (2014) as Cepheids and 178 stars -- as
type II Cepheids. We checked all these objects (classified as CEPH\_F,
CEPH\_1O, CEPH\_Other, T2CEPH\_N). The preliminary version of our
collection overlooked a surprisingly large number of 376 Cepheid candidates
from the EROS-2 catalog, of which 350 stars were monitored by the OGLE
survey (most of the remaining stars fell into the gaps between the CCD
detectors).

We carefully investigated the OGLE light curves of these sources and found
four additional classical Cepheids. Three of them were missed because of a
small number of observing points in the OGLE database and one was an
overtone Cepheid with the photometry affected by blending. We supplemented
our collection with these four Cepheids. The remaining 346 EROS-2
candidates for Cepheids turned out to be eclipsing binaries, ellipsoidal
variables, spotted stars, RR~Lyraes or other types of variable stars. This
result demonstrate the weakness of the automated methods of the variable
star classification.

We also confronted our collection with a list of 299 candidates for new
Cepheids in the SMC recently announced by the VMC near-infrared survey
(Moretti \etal 2015). We successfully identified OGLE-IV light curves for
278 of these stars and we confirm that 35 of them are real Cepheids (33
classical and two anomalous Cepheids). Most of the remaining objects from
the VMC list turned out to be constant or nearly constant stars.  Nine of
the Cepheids released by Moretti \etal (2015) were included in the GCVS
(Artyukhina \etal 1995). This example clearly indicates that only a full
variability survey with a sufficient number of observing epochs (at least
around one hundred) may provide data for reliable characterization and
classification of variable objects.

\Section{Discussion}
\Subsection{Period--Luminosity Relations}
The PL relations for Cepheids (also known as the Leavitt Law) were
discovered in the SMC (Leavitt 1908) and the Magellanic Clouds still play a
dominant role in the studies of this phenomenon (\eg Ngeow \etal 2009,
2010, 2012, Majaess \etal 2011, Storm \etal 2011, Macri \etal 2015). The PL
diagrams in the apparent {\it I} and {\it V} magnitudes and in the
extinction-free Wesenheit index, defined as $W_I=I-1.55(V-I)$, are shown in
\MakeTableee{c@{\hspace{15pt}}
           c@{\hspace{15pt}}
           r@{\hspace{15pt}}
           r@{\hspace{15pt}}
           r@{\hspace{9pt}}}{12.5cm}
{Period--Luminosity Relations for Classical Cepheids in the Magellanic Clouds}
{\hline
\noalign{\vskip3pt}
Mode of   & Galaxy & \multicolumn{3}{c}{$W_I=\alpha\log{P}+\beta$} \\
pulsation &        & \multicolumn{1}{c}{$\alpha$} & \multicolumn{1}{c}{$\beta$} & \multicolumn{1}{c}{$\sigma$} \\
\noalign{\vskip3pt}
\hline
\noalign{\vskip3pt}
F  & LMC & $-3.314\pm0.008$ & $15.888\pm0.005$ & $0.077$ \\
1O & LMC & $-3.431\pm0.007$ & $15.393\pm0.002$ & $0.081$ \\
2O & LMC & $-3.548\pm0.027$ & $15.025\pm0.008$ & $0.087$ \\
3O & LMC & $-4.000\pm0.134$ & $14.486\pm0.077$ & $0.071$ \\
\noalign{\vskip2pt}
\hline
\noalign{\vskip3pt}
F  & SMC & $-3.460\pm0.011$ & $16.493\pm0.005$ & $0.155$ \\
1O & SMC & $-3.548\pm0.017$ & $15.961\pm0.004$ & $0.169$ \\
2O & SMC & $-3.651\pm0.098$ & $15.545\pm0.025$ & $0.154$ \\
\noalign{\vskip15pt}
\hline
\noalign{\vskip3pt}
Mode of   & Galaxy & \multicolumn{3}{c}{$I=\alpha\log{P}+\beta$} \\
pulsation &        & \multicolumn{1}{c}{$\alpha$} & \multicolumn{1}{c}{$\beta$} & \multicolumn{1}{c}{$\sigma$}\\
\noalign{\vskip3pt}
\hline
\noalign{\vskip3pt}
F  & LMC & $-2.911\pm0.014$ & $16.822\pm0.009$ & $0.146$ \\
1O & LMC & $-3.260\pm0.013$ & $16.362\pm0.004$ & $0.162$ \\
2O & LMC & $-3.438\pm0.048$ & $15.940\pm0.014$ & $0.158$ \\
3O & LMC & $-3.829\pm0.197$ & $15.370\pm0.113$ & $0.104$ \\
\noalign{\vskip2pt}
\hline
\noalign{\vskip3pt}
F  & SMC & $-3.115\pm0.015$ & $17.401\pm0.007$ & $0.215$ \\
1O & SMC & $-3.299\pm0.023$ & $16.818\pm0.005$ & $0.222$ \\
2O & SMC & $-3.600\pm0.135$ & $16.350\pm0.034$ & $0.213$ \\
\noalign{\vskip15pt}
\hline
\noalign{\vskip3pt}
Mode of   & Galaxy & \multicolumn{3}{c}{$V=\alpha\log{P}+\beta$} \\
pulsation &        & \multicolumn{1}{c}{$\alpha$} & \multicolumn{1}{c}{$\beta$} & \multicolumn{1}{c}{$\sigma$} \\
\noalign{\vskip3pt}
\hline
\noalign{\vskip3pt}
F  & LMC & $-2.690\pm0.018$ & $17.438\pm0.012$ & $0.208$ \\
1O & LMC & $-3.142\pm0.018$ & $16.979\pm0.006$ & $0.227$ \\
2O & LMC & $-3.333\pm0.062$ & $16.531\pm0.018$ & $0.202$ \\
3O & LMC & $-3.719\pm0.257$ & $15.941\pm0.147$ & $0.135$ \\
\noalign{\vskip2pt}
\hline
\noalign{\vskip3pt}
F  & SMC & $-2.898\pm0.018$ & $17.984\pm0.008$ & $0.266$ \\
1O & SMC & $-3.155\pm0.028$ & $17.368\pm0.007$ & $0.271$ \\
2O & SMC & $-3.544\pm0.170$ & $16.870\pm0.043$ & $0.267$ \\
\noalign{\vskip3pt}
\hline}
\begin{figure}[p]
\hglue-3mm{\includegraphics[width=13cm]{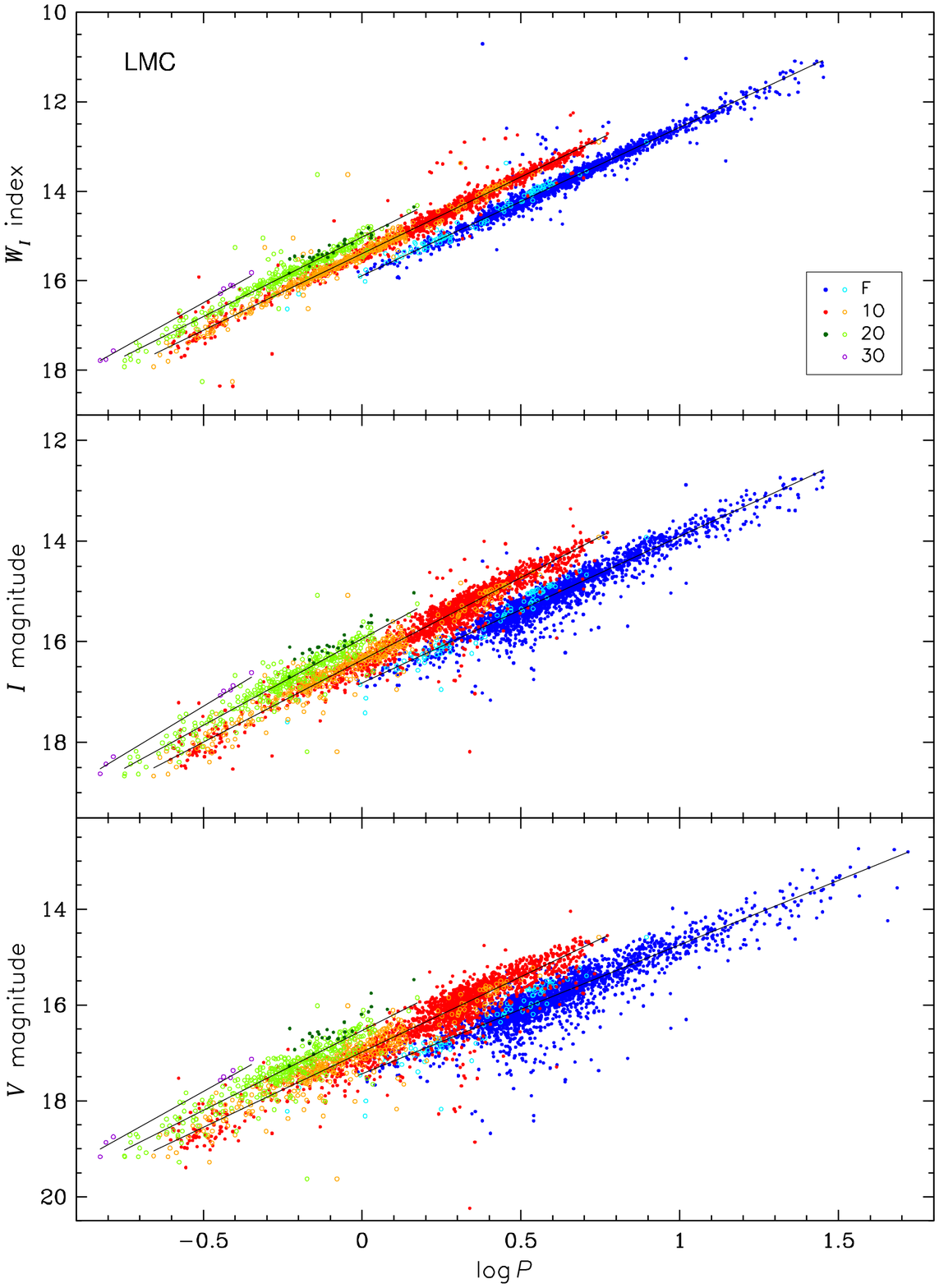}}
\vskip11pt
\FigCap{Period--luminosity diagrams for classical Cepheids in the
LMC. Blue, red, and dark-green solid circles mark F, 1O, and 2O single-mode
Cepheids, respectively. Cyan, orange, light-green and violet empty circles
represent, respectively, F, 1O, 2O, and 3O modes in multi-mode Cepheids.
Black lines show the linear least-square fits to the PL relations.}
\end{figure}
\begin{figure}[p]
\hglue-3mm{\includegraphics[width=13cm]{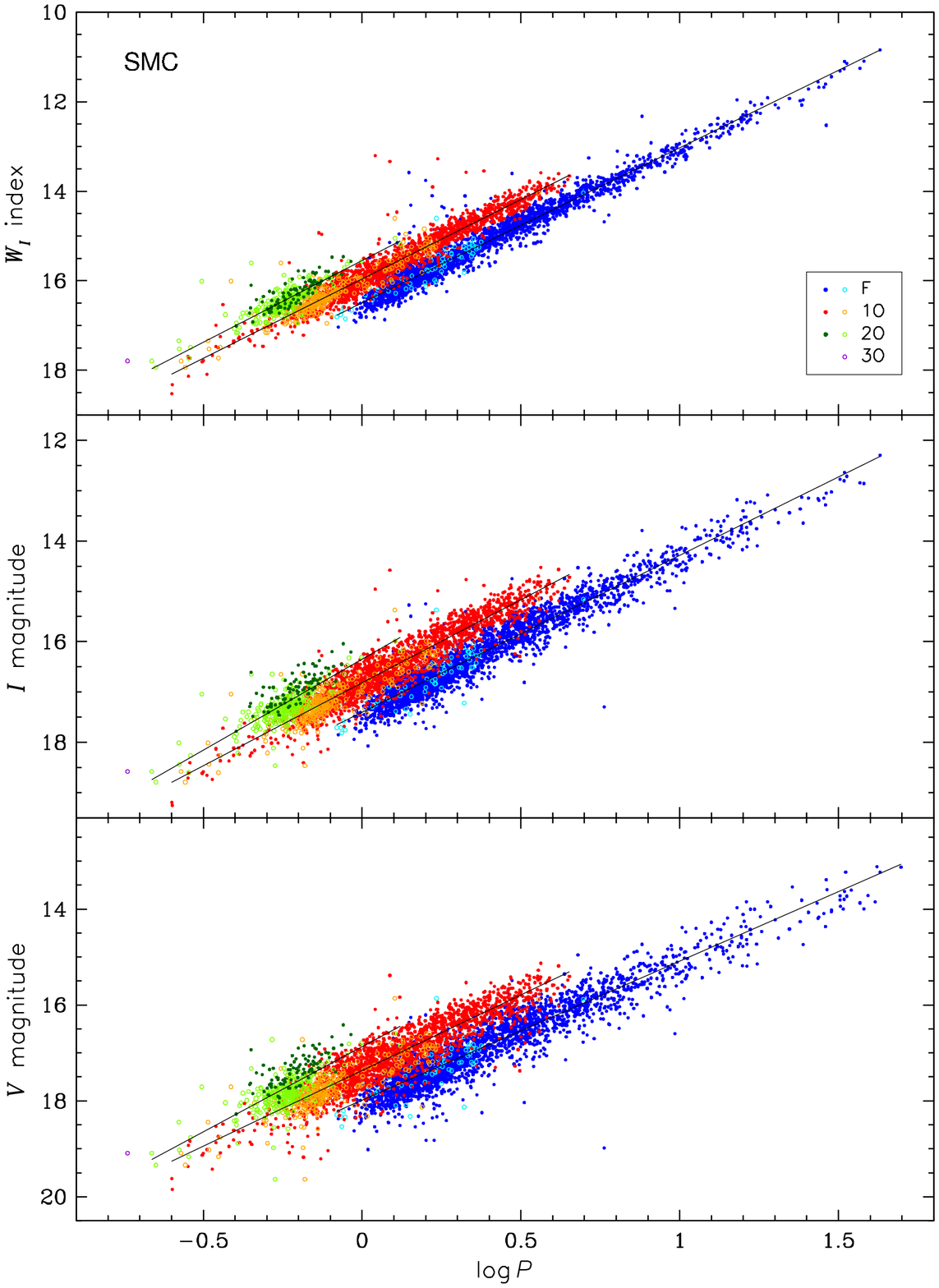}}
\vskip11pt
\FigCap{Period--luminosity diagrams for classical Cepheids in the
SMC. Color symbols represent the same modes of pulsation as in Fig.~2.}
\end{figure}
Figs.~2 and~3. In Table~2, we summarize the linear fits to the PL relations
presented in Figs.~2 and~3. We used the least square method with an
iterative $3\sigma$ clipping. In our procedure, we used all the Cepheids
pulsating in a given mode (including multi-mode pulsators), despite the
fact that some of the relations might not be strictly linear (\eg Ngeow
\etal 2009).

\begin{figure}[p]
\includegraphics[width=12.7cm]{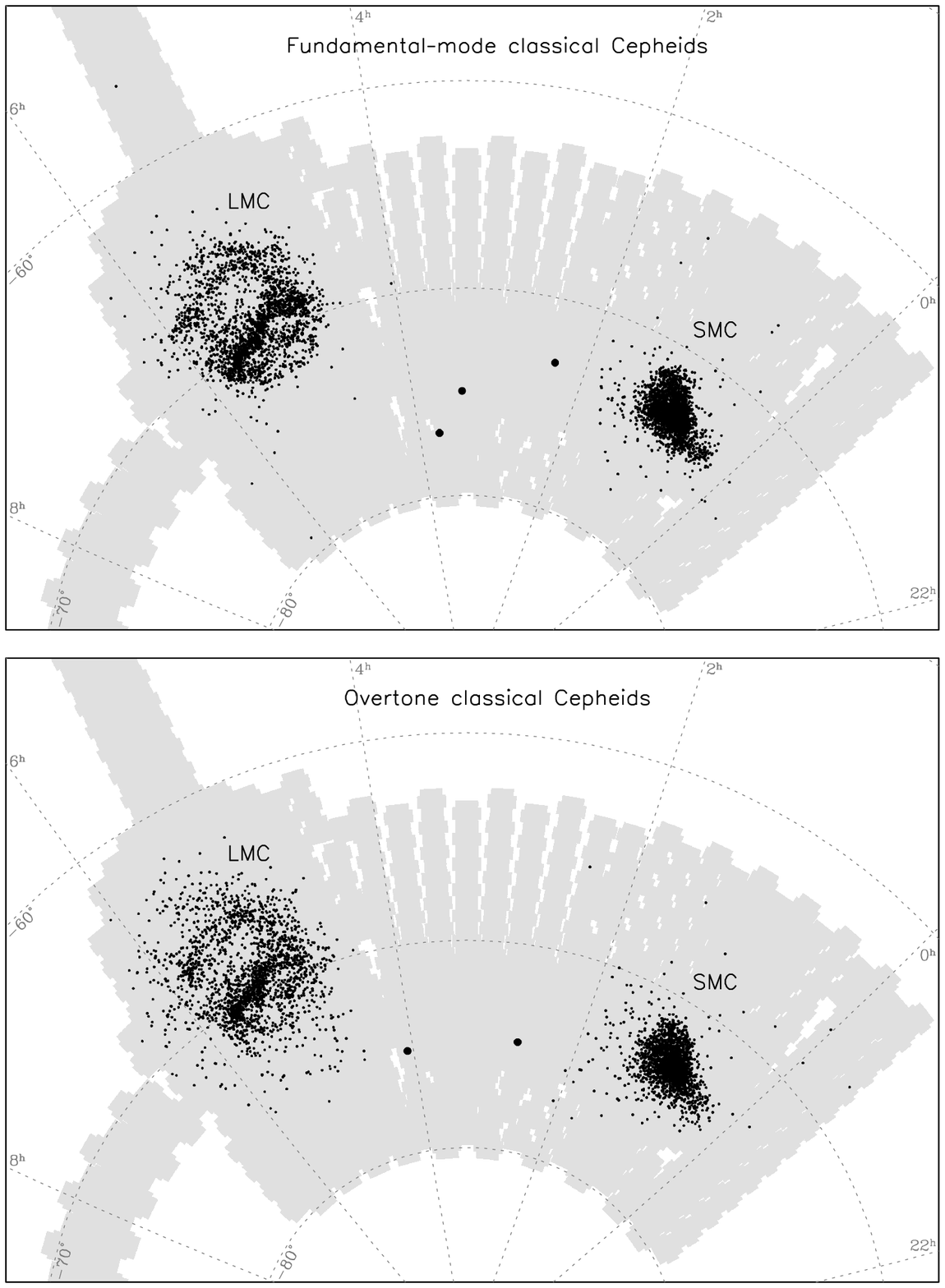}
\vspace*{5pt}
\FigCap{Spatial distribution of classical Cepheids in the Magellanic
System. {\it Upper panel} shows the positions of single-mode
fundamental-mode Cepheids. {\it Lower panel} presents single- and
multi-mode overtone pulsators. The gray area shows the sky coverage of the
OGLE fields. Larger dots mark classical Cepheids in the Magellanic Bridge.}
\end{figure}

\Subsection{Spatial Distribution of Classical Cepheids in the Magellanic
System}
Classical Cepheids are simultaneously distance indicators and tracers of
the young stellar population. Thus, they play a unique role in studying the
three-dimen\-sional structures of galaxies (\eg Haschke \etal 2012ab,
Moretti \etal 2014, Subramanian and Subramaniam 2015, Scowcroft \etal
2016). We believe that the OGLE-IV collection of $\delta$~Cep stars in the
Magellanic Clouds will be crucial for understanding the dynamical history
of both galaxies. In Fig.~4, we plot the two-dimensional distributions of
classical Cepheids in the Magellanic System, separately for the
fundamental-mode and first-overtone pulsators (together with the multi-mode
variables). It is clear that both classes of Cepheids have a different
spatial distribution, in particular in the LMC, which implies different
history of their formation.

Well defined and tight OGLE period--luminosity relations of classical
Cepheids derived in Section~6.1 provide an important tool for
the determination of distances to individual Cepheids, adding the third
dimension to the 2-D maps presented in Fig.~4. Such 3-D maps based on
complete sample of the OGLE collection of Cepheids will constitute a unique
picture of the Magellanic System enabling detailed modeling of their
structure as seen {\it via} young stellar population. Results of this study
will be presented in the forthcoming paper.

\Subsection{Classical Cepheids in the Outskirts of the Magellanic Clouds}
Several variable stars with Cepheid characteristics found in the OGLE-IV
data are located surprisingly far from the centers of the LMC and SMC in
the region called the Magellanic Bridge. This is a stream of gas and stars
distributed between these galaxies.

Careful analysis of the shape of the light curves of this sub-sample
indicated that some of these objects have characteristics similar to
anomalous Cepheids. Thus, they were included to the OGLE set of anomalous
Cepheids (Soszyñski \etal 2015). Their presence in the Magellanic
Bridge is not surprising because the distribution of anomalous Cepheids
clearly shows that these stars are often located in the halo of the
Magellanic Clouds, far from their centers (\cf Fig.~7, Soszyñski \etal
2015).

The remaining detected objects are genuine classical Cepheids. At least
five of them are located in the Magellanic Bridge. These are the first such
type variable stars discovered in this very important region of the
sky. The presence of classical Cepheids there independently confirms the
existence of young stellar component. This component of the Magellanic
Bridge was first discovered by Irwin \etal (1985, 1990) in a few regions
between the Magellanic Clouds. Recently, Skowron \etal (2014) presented
extensive density maps of various stellar populations in the entire
Magellanic Bridge based on the OGLE-IV observations. They showed that the
young stellar population forms a continuous stream connecting both
Magellanic Clouds.

\begin{figure}[htb]
\includegraphics[width=12.7cm]{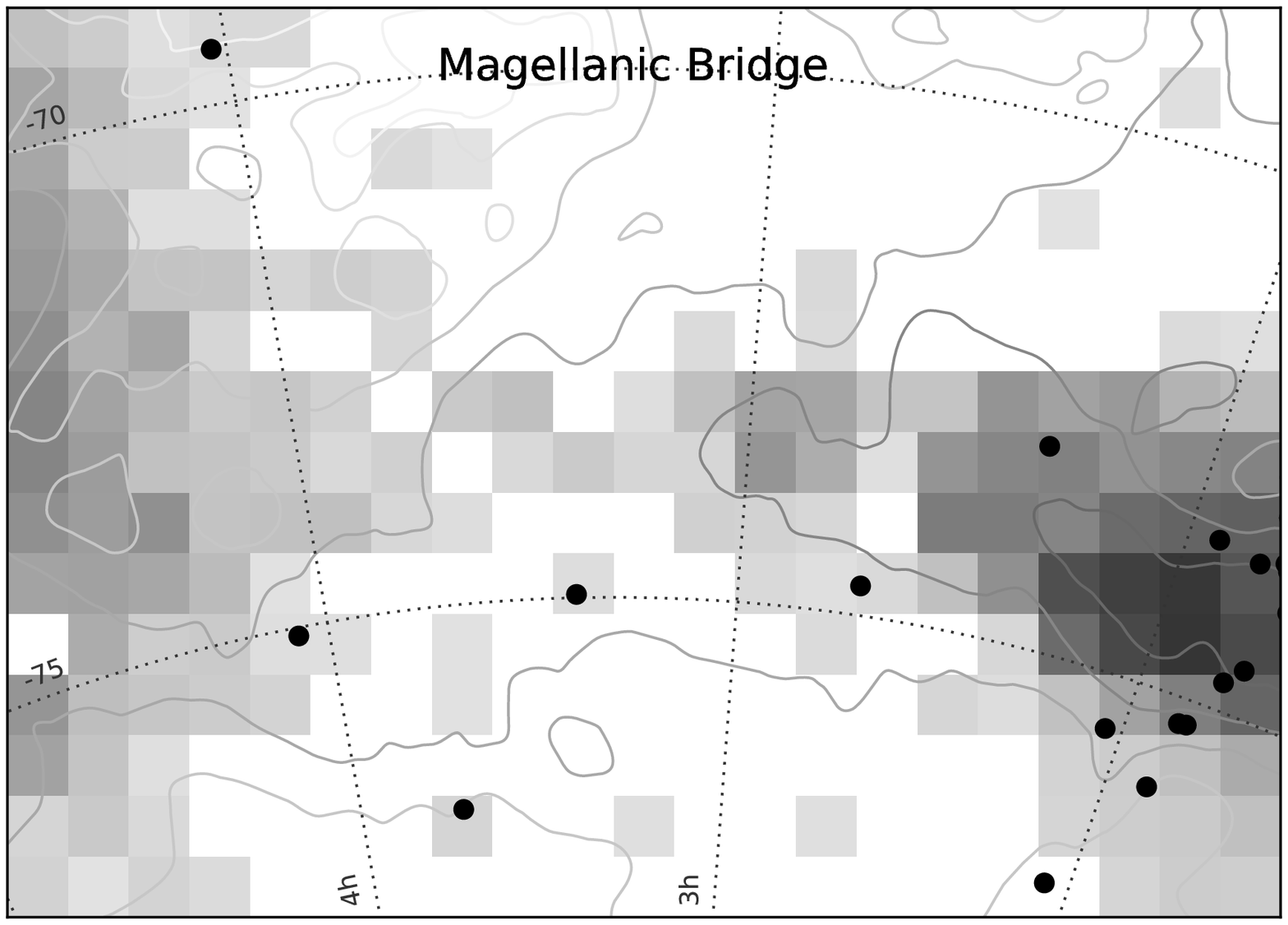}
\vspace*{5pt}
\FigCap{Spatial distribution of classical Cepheids in the Magellanic Bridge
area (black points). Background shaded map shows a spatial density map of
the young population stars from Skowron \etal (2014). Grey contours mark
neutral hydrogen emission from Kalberla \etal (2005).}
\end{figure}

Fig.~5 presents location of the Magellanic Bridge classical Cepheids on the
map of the distribution of the young stellar population in the Magellanic
Bridge. Positions of the five classical Cepheids in the Magellanic Bridge
coincide roughly with this stream although they are mostly located at its
southern side. We estimated the ages of these Cepheids using period--age
relations provided by Bono \etal (2005). The ages range from 27~Myr (for
OGLE-SMC-CEP-4953~=~HV~11211) to 280~Myr (for OGLE-SMC-CEP-4956). This
agrees with the scenarios of the past interaction between the Clouds which
took place about 250~Myr ago. Rough estimate of distances to the Magellanic
Bridge Cepheids based on their position in the LMC $W_I$ index PL diagram
indicates that objects located on the SMC side of the Bridge are at the
distance of the eastern wing of the SMC (55--60~kpc) while two objects
closer to the LMC are roughly at the LMC distance of
40--50~kpc. Surprisingly, two objects in the middle of the Magellanic
Bridge (OGLE-SMC-CEP-4956, OGLE-SMC-CEP-4957) are much farther than both
galaxies -- at a distance of 68--72~kpc. This indicates that the 3-D
distribution of matter in the Magellanic Bridge may be much more complex
than simple expectation from the recent encounter interaction of both
Clouds.

\Section{Conclusions}
We presented the most complete and least contaminated collection of
classical Cepheids in the Magellanic System. Compared to the past OGLE-III
Cepheid data release (Soszyñski \etal 2008, 2010), we increased the sizes
of our samples by 1249 and 327 Cepheids in the LMC and SMC,
respectively. Thus, we compiled a nearly complete census of these stars in
the Magellanic Clouds. We showed the distribution of classical Cepheids in
the Magellanic System and discovered the first classical Cepheids in the
Magellanic Bridge that confirm the presence and date the young stellar
population in this important region of the interaction between the
Magellanic Clouds.

The final OGLE collection of classical Cepheids should contribute to many
extensive astrophysical studies: better understanding of stellar evolution
and pulsation, modeling structure and dynamics of the Magellanic Clouds,
analysis of the extragalactic distance scale, and to many cosmological
applications.

\Acknow{We would like to thank Prof.\ M.~Kubiak, OGLE project co-founder,
  for his contribution to the collection of the OGLE photometric data over
  the past years. We are grateful to Z.~Ko³aczkowski and
  A.~Schwar\-zen\-berg-Czerny for providing software used in this study.

  This work has been supported by the Polish Ministry of Science and Higher
  Education through the program ``Ideas Plus'' award No. IdP2012
  000162. The OGLE project has received funding from the Polish National
  Science Centre grant MAESTRO no. 2014/14/A/ST9/00121 to AU.  DS is
  supported by the Polish National Science Center (NCN) under the grant
  no. 2013/11/D/ST9/03445.}

\end{document}